\documentclass[pre,aps]{revtex4}
\usepackage{graphicx,epsf}
\usepackage{amssymb}
\usepackage{amsmath}
\usepackage{latexsym}

\def\be{\begin{equation}}
\def\ee{\end{equation}}
\def\bq{\begin{eqnarray}}
\def\eq{\end{eqnarray}}
\def\beq{\begin{eqnarray*}}
\def\eeq{\end{eqnarray*}}

\begin{document}

\title{TORUS BREAKDOWN IN A UNI JUNCTION MEMRISTOR}

\author{Jean-Marc GINOUX$^{1}$, Riccardo MEUCCI$^{2}$ Stefano EUZZOR$^{2}$, Angelo DI GARBO$^{3}$}

\affiliation{$^1$Laboratoire LIS, CNRS, UMR 7020, Universit\'{e} de Toulon, BP 20132, F-83957 La Garde cedex, France}

\affiliation{$^2$Istituto Nazionale di Ottica, Consiglio Nazionale delle Ricerche, Largo E. Fermi 6, 50125 Firenze, Italy}

\affiliation{$^3$Istituto di Biofisica, Consiglio Nazionale delle Ricerche, Via G. Moruzzi 1, 56124 Pisa, Italy}

\begin{abstract}
Experimental study of a uni junction transistor (UJT) has enabled to show that this electronic component has the same features as the so-called ``memristor''. So, we have used the memristor's direct current (DC) $v_{M}-i_{M}$ characteristic for modeling the UJT's DC current-voltage characteristic. This led us to confirm on the one hand, that the UJT is a memristor and, on the other hand to propose a new four-dimensional autonomous dynamical system allowing to describe experimentally observed phenomena such as the transition from a limit cycle to torus breakdown.
\end{abstract}

\maketitle

\section{From Uni Junction Transistor to Uni Junction Memristor}

\subsection{The Uni Junction Transistor}

The Uni Junction Transistor (UJT) was discovered by accident by Jerry Suran in 1953. At that time he was working in the circuits group under Richard Shea and Arnold Lesk evaluating experimental tetrode transistors made by John Saby's group. In an interview given half a century later to \cite{Ward2005}, Suran explained that he asked to Saby:

\begin{quote}
``\ldots if he could maybe build some tetrodes for us to experiment with.  Just try to put two ohmic contacts across the base of a transistor, and try to build the transistor just the way he would normally build a triode.  We were just curious to see how these things would work. I believe that this particular experiment was run about a year after I came to the Electronics Laboratory, so it was probably about the middle of 1953, or maybe a little bit later in that year. What happened was that the first couple of tetrodes that we got we could notice very little effect of the electric field.  Our theory just wasn't born out, that this fourth electrode was going to do anything at all.  On the other hand, one of those tetrodes curiously had a hysteresis effect on the input, and when we put an oscilloscope on it we found that the thing was oscillating.  There was no effect of input voltage or current on the output, so it became apparent quickly that something had happened to the collector contact and that this one had a broken lead.''
\end{quote}

\begin{figure}[htbp]
\centerline{\includegraphics[width = 2.5cm, height = 7.11cm]{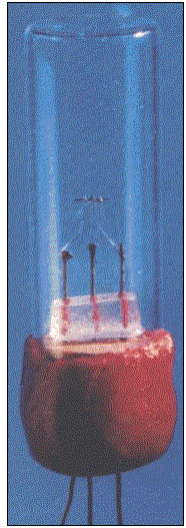}}
\caption{Original double base diode made by Arnie Lesk in John Saby's lab in 1953, from \textit{Electronics} (Feb. 19, 1968).}
\label{fig1}
\end{figure}

\begin{quote}
``Remember these germanium transistors in those days, the experimental ones that we were building in the laboratory, were put in a little vial, a tiny test tube filled with silicone oil to stabilize the surface.  Then it was sealed with a wax seal on the input where the wires came into the transistor.  The wax seal was just to stabilize the wires.   So, we were experimenting with devices that were built that way, and I guess a collector lead had broken off, and in a place we couldn't see it. That's how we got the name, ``double base diode''.  When you measure the dc characteristics of the input, when you tie the two bases together, it was just like a diode, but when you put a field across the two base contacts, this thing oscillated, and it was oscillating with the parasitic capacitance of our instrumentation on the input side. So, we were quickly then able to determine that we had a device that was very different from what we expected, and it was a serendipity effect because the collector contact had opened up quite unexpectedly - as far as I can recall, that was how the unijunction transistor was discovered.''
\end{quote}

Thus, the UJT also called ``double base diode'' has three terminals: an emitter ($E$) and two bases ($B_1$ and $B_2$) (see Fig. 2).

\begin{figure}[htbp]
\centerline{\includegraphics[width = 7.15cm, height = 4cm]{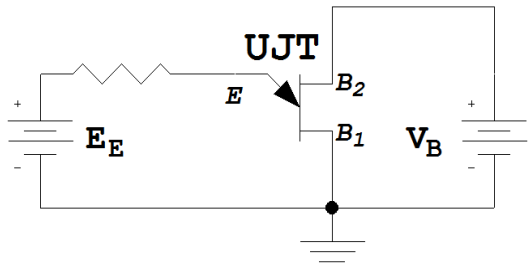}}
\caption{Schematic structure of the UJT with the bias voltage.}
\label{fig2}
\end{figure}

The UJT has been mainly used as active device in relaxation oscillators because its current-voltage \textit{static characteristics} has a portion in which the resistance is negative at the emitter terminal (E) (see Fig. 3). The static emitter characteristic, i.e., the mathematical function modeling the relationship between the emitter voltage $V_E$ and the emitter current $I_E$ of a UJT at a given inter base voltage $V_{B}$ is plotted in Figure 3.

\begin{figure}[htbp]
\centerline{\includegraphics[width = 10.31cm, height = 7cm]{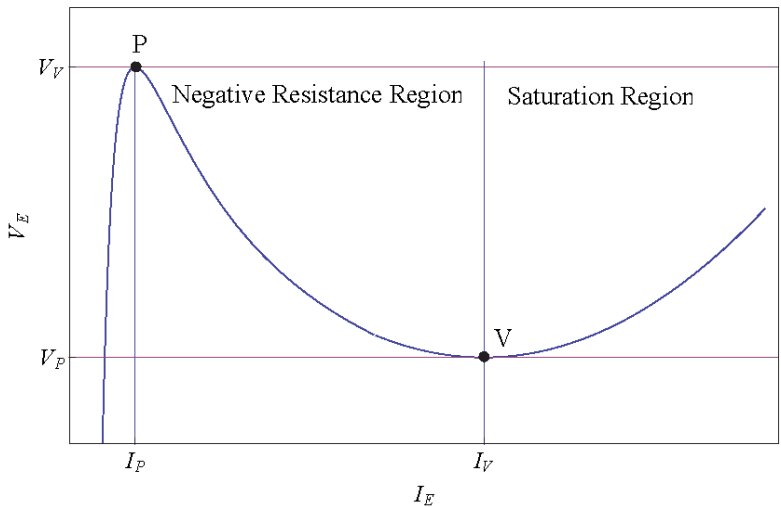}}
\caption{Static Emitter Characteristic of the UJT.}
\label{fig3}
\end{figure}

From the static emitter characteristic of the UJT plotted in Fig. 3, we notice that $V_E$ increases rapidly till it reaches a maximum value at the peak point $P$. In the meantime the emitter current $I_E$ is almost zero. This portion of the curve corresponds to a region called the ``cut-off'' region. Once conduction is established at $V_E = V_P$ the emitter potential $V_E$ starts decreasing drastically with the rapid increase in emitter current $I_E$ till it reaches a maximum value at the valley point $V$. As a result a negative resistance portion appears in the characteristic curve. Beyond the valley point, any further increase in the emitter current $I_E$ places the UJT in the saturation region.

\subsection{The memristor}

On April 30$^{th}$ 2008, the journal \textit{Nature} announced that the missing circuit element, postulated thirty-seven years before by Professor Leon O. Chua has been found \cite{Strukhov2008}. However, contrary to what one might think, it is not by experimenting, but by logical deduction that Professor L. O. Chua was able to postulate the existence of a missing circuit element. In his now famous publication of 1971, \citet{Chua1971} considered the three basic building blocks of an electric circuit: the capacitor, the resistor and the inductor as well as the three laws linking the four fundamental circuit variables, namely, the electric \textit{current} $i$, the \textit{voltage} $v$, the \textit{charge} $q$ and the \textit{magnetic flux} $\varphi$ (see Fig. 4).

\begin{figure}[htbp]
\centerline{\includegraphics[width=6.66cm,height=6.5cm]{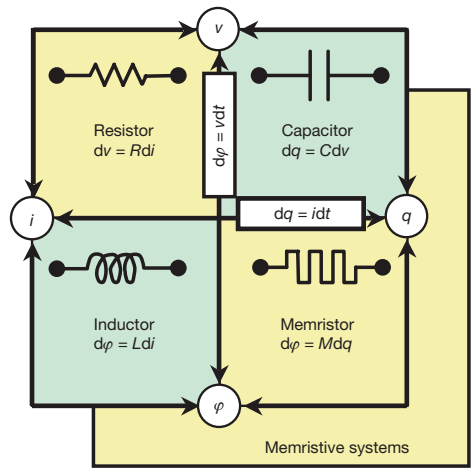}}
\caption{The four fundamental two-terminal circuit elements, \cite[p. 80]{Strukhov2008}}
\label{fig4}
\end{figure}

Then, \citet[p. 507]{Chua1971} explained that:

\begin{quote}

``\ldots by the \textit{axiomatic} definition of the three classical circuits elements, namely, the \textit{resistor} (defined by a relationship between $v$ and $i$), the \textit{inductor} (defined by a relationship between $\varphi$ and $i$), and the \textit{capacitor} (defined by a relationship between $q$ and $v$). Only one relationship remains undefined, the relationship between $\varphi$ and $q$.''

\end{quote}

He thus concluded from the logical as well as axiomatic points of view, that it is necessary, for the sake of \textit{completeness}, to postulate the existence of a fourth circuit element to which he gave the name \textit{memristor} since it behaves like a nonlinear resistor with memory \cite{Ginoux2013}. Unlike the transistor that allows the current to flow or not, and so uses bits (0 or 1), the memristor has a variable resistance and can therefore take all the values between 0 and 1. Depending on the incoming signal and its previous state, the memristor adjusts its resistance to current and keep in memory its previous state, hence its name.

\begin{figure}[htbp]
\centerline{\includegraphics[width=7cm,height=6.5cm]{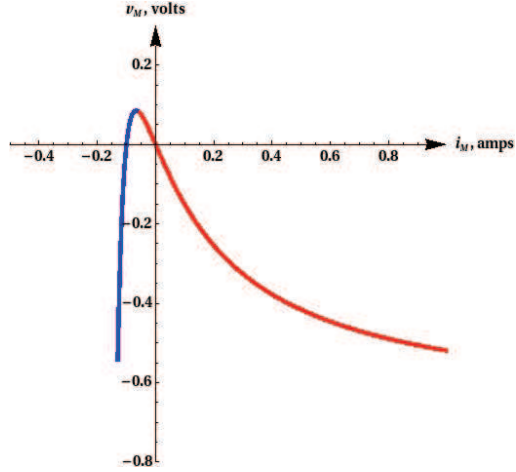}}
\caption{The DC $v_M-i_M$ static characteristic of the memristor, \cite[p. 1574]{Muthuswamy2010}.}
\label{fig5}
\end{figure}

Eight years ago, \citet[p. 1574]{Muthuswamy2010} plotted the direct current (DC) $v_{M}-i_{M}$ static characteristic of the memristor (see Fig. \ref{fig5}).

In their paper, \citet[p. 1580]{Muthuswamy2010} stated the equation of this curve representing the direct current (DC) $v_{M}-i_{M}$ static characteristic of the memristor:

\begin{equation}
v_M =i_M\left( -1 + \frac{i_M^2}{\left( i_M + \alpha \right)^2}\right)\beta
\label{eq1}
\end{equation}

By comparing the static characteristic of the UJT plotted in Fig. \ref{fig3} with that of the memristor plotted in Fig. \ref{fig5}, it appears that both curves shape is similar even identical. Hence the idea to use the memristor's direct current (DC) $v_{M}-i_{M}$ characteristic for modeling the UJT's DC current-voltage characteristic. Although fundamental differences exist between transistor and memristor especially the capability of this latter of altering its resistance and storing multiple memory states \cite{Statho2017}, searchers still consider memristor as an hybrid, halfway between a transistor and a memory. In a previous publication, we have studied the complex dynamics of a direct current glow discharge tube, i.e. a neon tube \cite{Pugliese2015}. This led us to propose a new four-dimensional autonomous dynamical system for which we have highlighted bifurcations routes from torus breakdown to homoclinic chaos following the Newhouse-Ruelle-Takens scenario \cite{Ginoux2017}. Starting from an electronic circuit comprising a UJT associated with an oscillating circuit, we show in this work that the memristor's characteristic (\ref{eq1}) can be used for modeling that of the UJT so, that's why we suggest to call it ``Uni Junction Memristor''. Then, while using Kirchoff's law, we derive the ordinary differential equations representing the various oscillatory regimes of a UJT. As previously, mathematical analysis and detailed numerical investigations of the four-dimensional dynamical system thus built, enable to confirm the torus breakdown observed experimentally.

\section{Experimental Setup}

In our laboratory, we have built an electronic circuit comprising a UJT. A schematic representation of the experimental setup is plotted in Fig. 6, where the UJT has been replaced by an equivalent model including a fixed resistor $R_{b2}$, a variable and current dependent resistor $R_{b1}$ and an inductor $L$ accounting for parasitic effects.

\begin{figure}[htbp]
\centerline{\includegraphics[width=14.41cm,height=8.68cm]{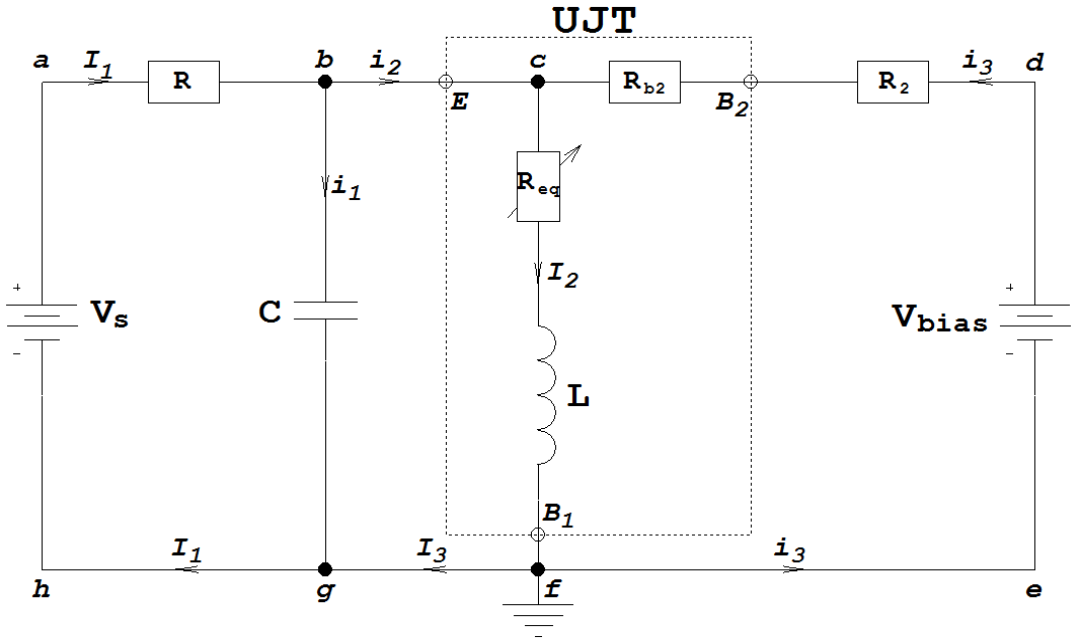}}
\caption{Circuit diagram of the experimental setup.}
\label{fig6}
\end{figure}

By using Kirchhoff's laws, it is not difficult to show that:

\begin{equation}
I_3 = i_2 \quad \mbox{and} \quad I_1 = i_1 + i_2
\label{eq2}
\end{equation}

\begin{equation}
\label{eq3}
i_1 = C\dfrac{dV_{UJT}}{dt},
\end{equation}

By considering the circuit loop (abcfgh) in Fig. \ref{fig6}, we obtain:

\begin{equation}
\label{eq4}
i_1 + i_2 = \dfrac{V_S - V_{UJT}}{R},
\end{equation}

Equations (\ref{eq2}-\ref{eq4}) lead to

\begin{equation}
\label{eq5}
\dfrac{dV_{UJT}}{dt} = \dfrac{1}{RC} \left( V_S - V_{UJT} - R i_2 \right).
\end{equation}

Since the potential difference $V_{cf} = V_{UJT}$, we have:

\begin{equation}
\label{eq6}
L\dfrac{d(i_2+i_3)}{dt} = V_{UJT} - R_{eq} \left( i_2 + i_3 \right),
\end{equation}

where $L$ represents the spurious inductance of the UJT. Then, by considering the circuit loop (dcfe), we find that

\begin{equation}
\label{eq7}
i_3 = \dfrac{V_{bias} - V_{UJT}}{R_2 + R_{b2}},
\end{equation}

Let's pose $i = i_2 + i_3$ and let's replace $i_3$ by the right hand side of Eq. (\ref{eq7}) in Eq. (\ref{eq5}) and (\ref{eq6}) which read:

\begin{equation}
\label{eq8}
\dfrac{dV_{UJT}}{dt} = \dfrac{1}{RC} \left[ \left( V_S  + \dfrac{RV_{bias}}{R_2 + R_{b2}} \right)- \left(1 + \dfrac{R}{R_2 + R_{b2}} \right) V_{UJT} - R i  \right],
\end{equation}

\begin{equation}
\label{eq9}
L\dfrac{di}{dt} = V_{UJT} - R_{eq} i.
\end{equation}

Let's notice that $R_{eq} i$ represents the nonlinear electromotive force of the UJT which can be modeled from its current-voltage characteristics by a function $G(i) = R_{eq} i$. Then, we introduce ``an oscillatory modulation around the working point to reproduce the experimental time series'' by coupling the following equation of the damped oscillator with Eqs. (\ref{eq8} \& \ref{eq9}):

\begin{equation}
\label{eq10}
\ddot{z} + \beta \dot{z} + \omega^2 z = \gamma x.
\end{equation}

and by posing:

\[
V_{bias} \to V_{bias} + m z \quad \mbox{with} \quad 0 \leqslant x \leqslant 1.
\]

where $m$ will play the role of control parameter. Thus, taking into account this ``modulation effect'', Eq. (\ref{eq8}) reads now

\[
\dfrac{dV_{UJT}}{dt} = \dfrac{1}{RC} \left[ \left( V_S  + \dfrac{RV_{bias}}{R_2 + R_{b2}} \right) + \dfrac{R}{R_2 + R_{b2}} m z - \left(1 + \dfrac{R}{R_2 + R_{b2}} \right) V_{UJT} - R i  \right]
\]

In order to simplify the Eqs. (\ref{eq8} \& \ref{eq9}), let's pose

\[
\begin{aligned}
V'_{bias} & = V_S  + \dfrac{RV_{bias}}{R_2 + R_{b2}}, \hfill \\
k & = 1 + \dfrac{R}{R_2 + R_{b2}}, \hfill \\
\end{aligned}
\]

where $V'_{bias}$ is in Volt and $k$ is a dimensionless parameter. So, we have:

\[
\dfrac{dV_{UJT}}{dt} = \dfrac{1}{RC} \left[ V'_{bias} + \left( k - 1 \right) m z  - k V_{UJT} - R i  \right]
\]

The experimental current-voltage characteristic of the UJT reaches an asymptotic value of about $v = 1$ V while the magnitude of current $i$ is of order of magnitude $10^{-3}$ A. So, we will use these values to rescale and dedimensionalize Eqs. (\ref{eq8} \& \ref{eq9}). Thus, we pose $V_{UJT} = vy/k$ with $v = 1$ V and $i = \alpha k x$ with $\alpha = 10^{-3}$ A. We use also a dimensionless time variable $t \to \beta_1 t$ where $\beta_1 = 10^{-4}$ s and we introduce the following new parameters set:

\[
A_0 = \dfrac{\beta_1}{\beta_2}V'_{bias}, A_1 = \dfrac{\beta_1}{\beta_2}, A_2 = A_1 R \alpha k.
\]

with $\beta_2 = RC/k$ (let's remind that [RC] = s). Eqs. (\ref{eq8} \& \ref{eq9}) read then:

\begin{equation}
\label{eq11}
\dfrac{dy}{dt} = A_0 - A_1 y - A_2 x + A_1 \left( k - 1 \right)mz,
\end{equation}

\begin{equation}
\label{eq12}
\mu \dfrac{dx}{dt} = y - g\left( x \right).
\end{equation}

where we have posed:

\[
\mu = \dfrac{\alpha L k^2}{\beta_1 v} \quad \mbox{and} \quad G\left( i \right) = \dfrac{v}{k} g\left( x \right).
\]

Let's notice that Eq. (\ref{eq10}) can be written as a set of first order ordinary differential equations:

\[
\begin{aligned}
\dfrac{dz}{dt} & = u, \hfill \\
\dfrac{du}{dt} & = - \beta u - \omega^2 z + \gamma x.
\end{aligned}
\]

Thus, by posing $z \to z /v(k - 1)$ and $u \to u / \beta_1 v (k-1)$, we obtain:

\[
\begin{aligned}
\dfrac{dz}{dt} & = u, \hfill \\
\dfrac{du}{dt} & = - \beta \beta_1 u - \left(\beta_1 \omega \right)^2 z + \dfrac{\gamma \left( k - 1 \right)\beta_1^2}{v} x.
\end{aligned}
\]

Since $\beta$, $\omega$ and $\gamma$ are free parameters we can pose to simplify: $\beta \beta_1 \to \beta$, $\beta_1 \omega \to \omega$, $\gamma \left( k - 1 \right)\beta_1^2 /v \to \gamma$. \\

Finally, the UJT circuit model reads:

\begin{equation}
\label{eq13}
\begin{aligned}
\mu \dfrac{dx}{dt} & = y - g\left( x \right), \hfill \\
\dfrac{dy}{dt} & = A_0 - A_1 y - A_2 x + A_1 m z, \hfill \\
\dfrac{dz}{dt} & = u, \hfill \\
\dfrac{du}{dt} & = - \beta u - \omega^2 z + \gamma x.
\end{aligned}
\end{equation}

So, according to what precedes (see Sect. 1), we propose to use the memristor's direct current (DC) $v_{M}-i_{M}$ characteristic (\ref{eq1}) for modeling the UJT's DC current-voltage characteristic $g(x)$. Hence we have the dimensionless form following function:

\begin{equation}
\label{eq14}
g\left( x \right) = x\left( a + \dfrac{bx^2}{\left( x +c \right)^2} \right)
\end{equation}

The parameters values $a$, $b$ and $c$ obtained by fitting the experimental data provides:

\[
a = 419.888, \quad b = 422.443, \quad c = 0.0129707,\\
\]

with the coefficient of determination $R^2 = 0.990896$ indicating a quite good fit of the data \cite{Bjorck1996}. In our experimental setup, the electrical components in the circuit diagram plotted in Fig. 6 have the following values: $V_S = 7 V$, $V_{bias} = 4.67 V$, $R = 12.6 k \Omega$, $R_2 = 677 \Omega$, $R_{b2} = 2.5 k \Omega$, $C = 49.73 nF$ and $L = 0.2 mH$. Thus, with these values parameters set is:

\[
\mu = 0.05, \mbox{ } A_0 = 26, \mbox{ } A_1 = 0.7925, \mbox{ } A_2 = 49.59, \mbox{ } \beta = 0.008, \mbox{ } \omega = 1.2 \mbox{ and } \nu =1.
\]

\section{Stability Analysis}

\subsection{Fixed points}

Fixed points are determined while using the classical nullclines method. Thus, by plugging $y$ and $z$ in the first equation of the dynamical system (\ref{eq13}) leads to the unique positive fixed point of this four-dimensional dynamical system which reads:

\begin{equation}
\label{eq14}
I \left(x^*, y^* = \dfrac{A_0}{A_1} - \dfrac{A_2}{A_1}x^* + m \dfrac{\gamma}{\omega^2}x^*, z^* = \dfrac{\gamma}{\omega^2}x^*, u^* = 0 \right),
\end{equation}

where the expression of $x^*$ (too large to be explicitly written here) depends on the control parameter $m$.\\

\subsection{Jacobian matrix}

The Jacobian matrix of dynamical system (\ref{eq13}) reads:

\begin{equation}
\label{eq15}
J = \begin{pmatrix}
- \dfrac{1}{\mu} \dfrac{a (c+x)^3 + b x^2 (3c + x)}{(c+x)^3} \  &  \  \dfrac{1}{\mu} \  &  \  0 \  &  \  0 \vspace{6pt} \\
-A_2 \  &  \  -A_1 \  &  \  m A_1 \  &  \  0 \vspace{6pt} \\
0 \  &  \  0 \  &  \  0 \  &  \  1 \vspace{6pt} \\
\gamma \  &  \  0 \  &  \  - \omega^2 \  &  \  - \beta \vspace{6pt}
\end{pmatrix}
\end{equation}

By replacing the coordinate of the fixed point $I$ (\ref{eq14}) in the Jacobian matrix (\ref{eq15}) one obtains the Cayley-Hamilton fourth degree eigenpolynomial from which one can deduce the two pair of two complex conjugate eigenvalues $\lambda_i$ depending on $m$. The real part of the first pair is strictly negative while the second one is strictly positive for all values of $m \in [0,1]$. So, the fixed point $I$ is unstable according to Lyapunov's theorem \cite{lyapunov1892} and no Hopf bifurcation could occurred with this parameter set.

\subsection{Bifurcation diagram}

Thus, in order to highlight the effects of the control parameter $m$ changes on the topology of the attractor we have  built a bifurcation diagram (see Fig. 7) that we have compared to the phase portraits plotted in Fig. 8. We observe that for $m = 0$, the attractor is a \textit{limit cycle} (see Fig. 8a). As $m$ increases between $0$ and $0.35$, the \textit{limit cycle} thickens into the $u$-direction and becomes a``ring'' consisting in dense trajectories (see Fig. 8b). For $m \approx 0.35$, the trajectories split into two parts and form a kind of ``spring'' (see Fig. 8c). In the interval $0.35 \leqslant m \leqslant 0.7$, the particular feature of the bifurcation diagram shows that the trajectories are not dense on the attractor. Then, for $m \approx 0.5$, a ``torus'' appears (see Fig. 8d).

\newpage

\begin{figure}[htbp]
\centerline{\includegraphics[width=12.22cm,height=8cm]{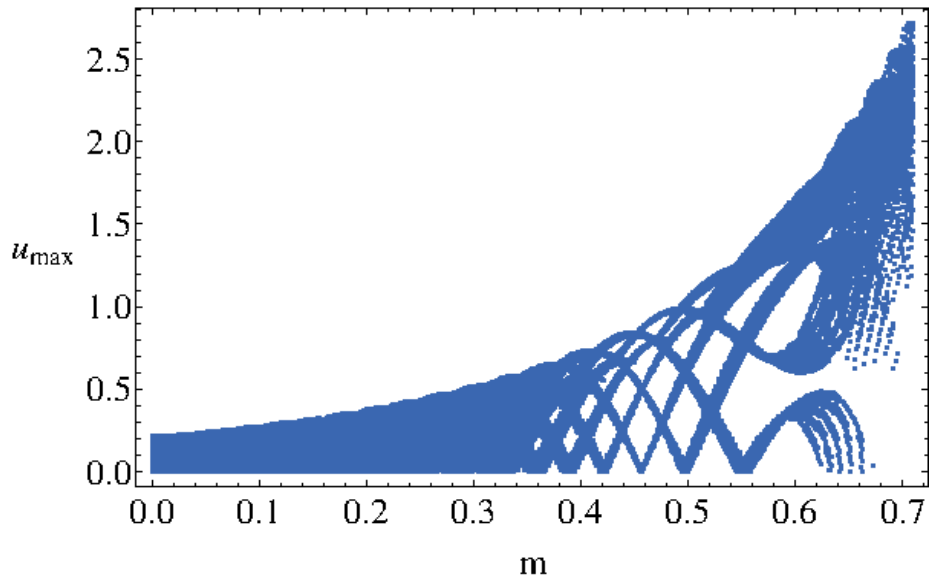}}
\caption{Bifurcation diagram $u_{max}$ as function of $m$.}
\label{fig7}
\end{figure}

\begin{figure}[htbp]
  \begin{center}
    \begin{tabular}{ccc}
      \includegraphics[width=7.05cm,height=7.05cm]{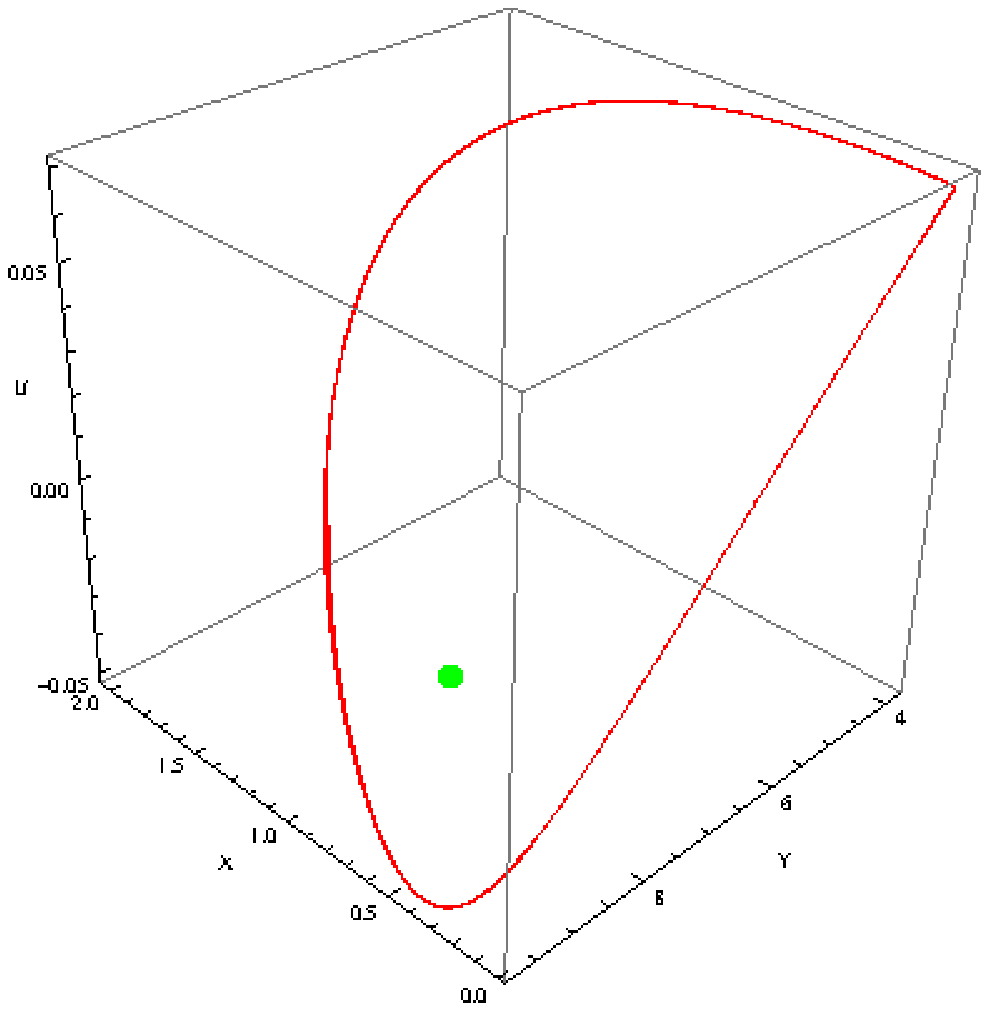} & ~~~~~~ &
      \includegraphics[width=7.05cm,height=7.05cm]{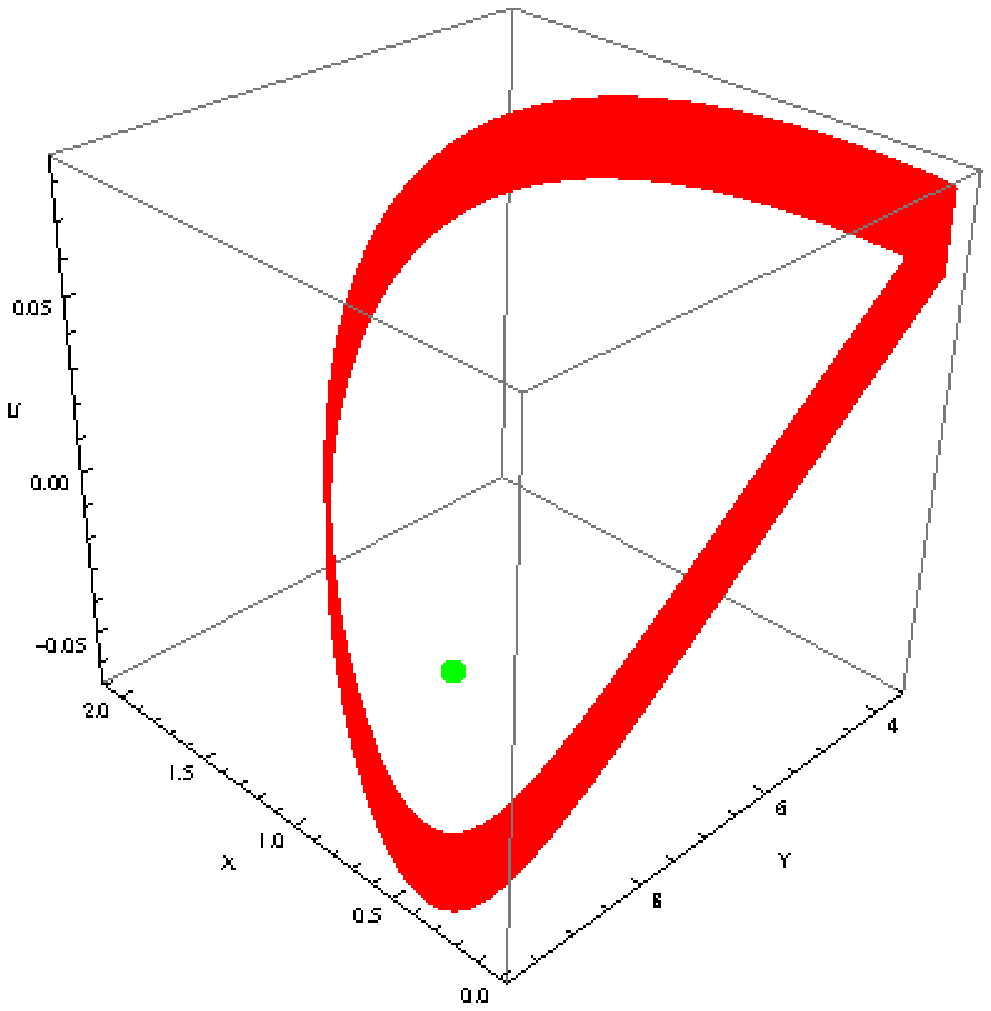} \vspace{0.1in} \\
      (a) $m = 0$ & & (b) $m = 0.3$ \\[0.2cm]
      \includegraphics[width=7.05cm,height=7.05cm]{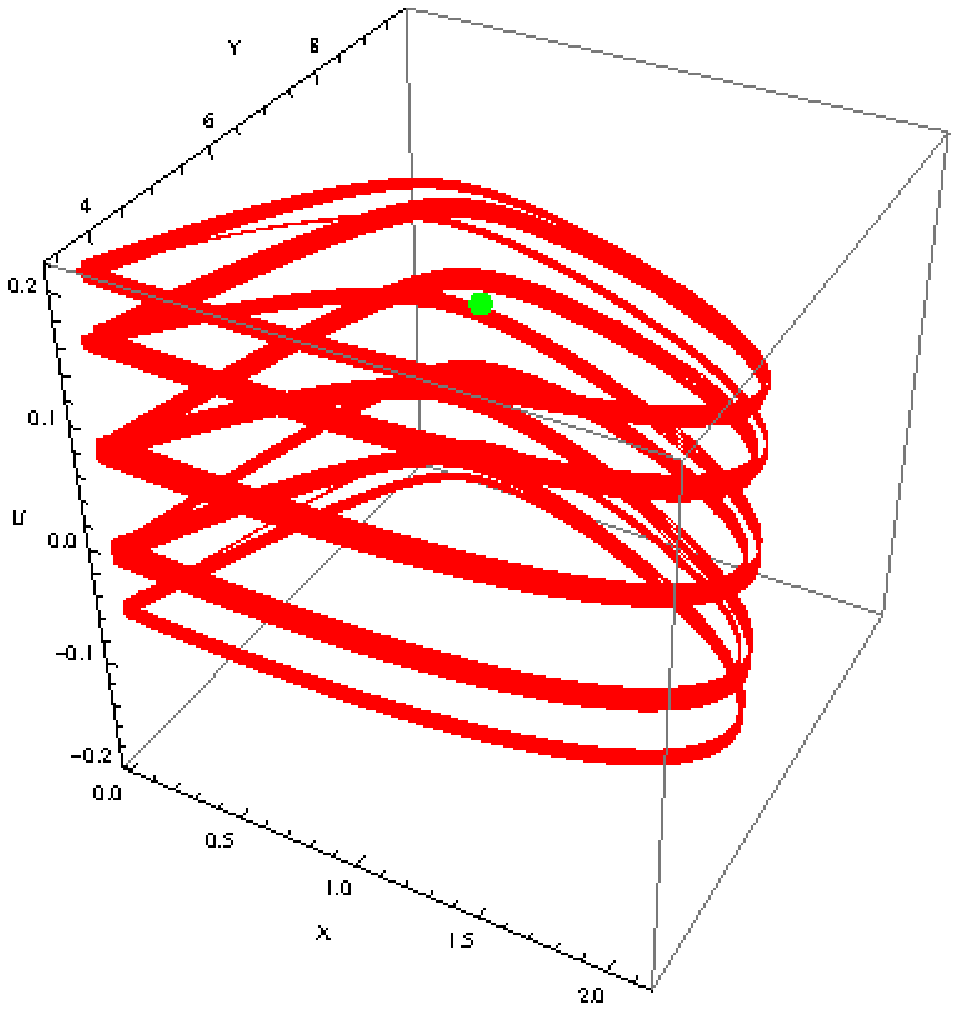} & ~~~ &
      \includegraphics[width=7.05cm,height=7.05cm]{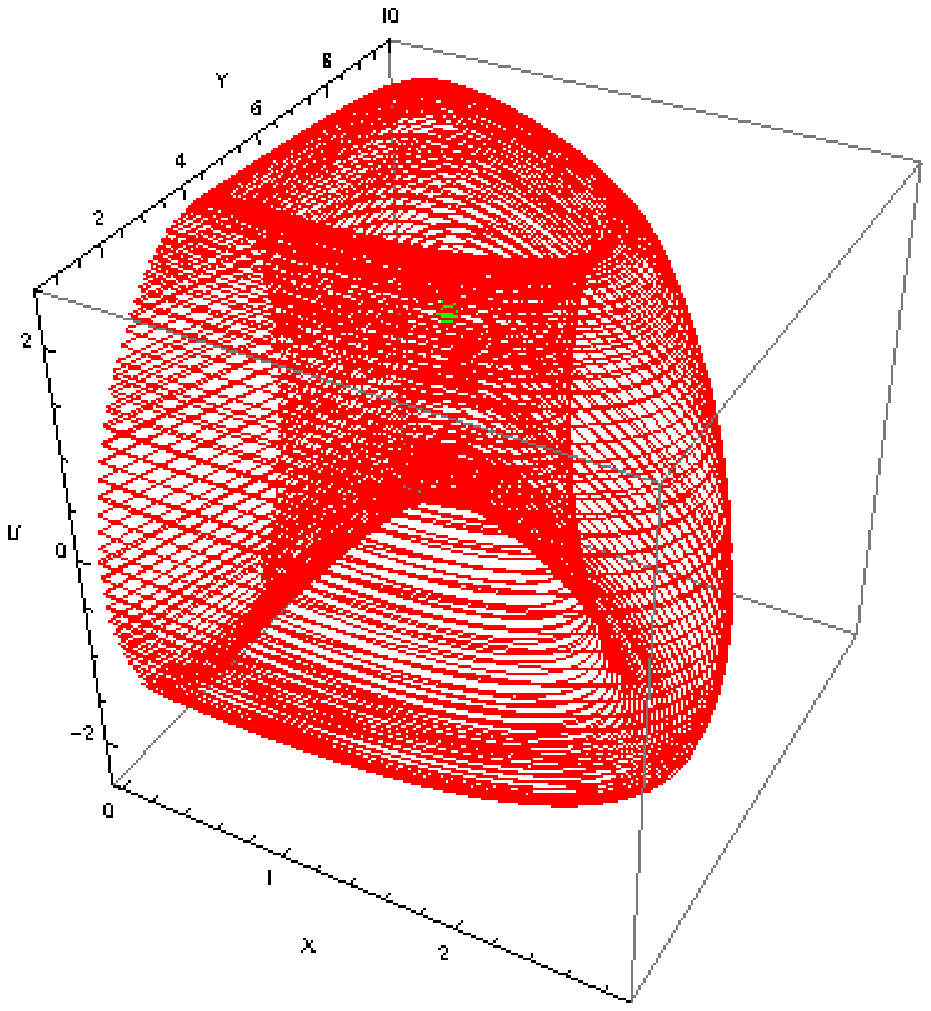} \vspace{0.1in} \\
      (c) $m = 0.4$ & & (d) $m = 0.5$ \\[-0.2cm]
    \end{tabular}
    \caption{Phase portraits of dynamical system (\ref{eq13}) in the ($x,y,u$)-space for various values $m$.}
    \label{fig8}
  \end{center}
  \vspace{-0.5cm}
\end{figure}

By reading from right to left the bifurcation diagram presented in Fig. \ref{fig7} and comparing it to the phase portraits plotted in Fig. \ref{fig8}, topological changes of the trajectory curve, integral of dynamical system (\ref{eq13}) seem to follow these transitions:

\[
\mbox{Torus} \Rightarrow \mbox{Chaos} \Rightarrow  \mbox{Limite Cycle}
\]

In order to confirm such scenario, Lyapunov Characteristic Exponents (LCE) have been computed in each case.

\subsection{Numerical computation of the Lyapunov exponents}

The algorithm developed by Pr. Marco Sandri [1996] for Mathematica$^{\mbox{\scriptsize{\textregistered}}}$ has been used to perform the numerical calculation of the Lyapunov characteristics exponents (LCE) of dynamical system (\ref{eq13}) in each case. LCEs values have been computed within each considered interval ($m \in [0, 0.5]$ and $[0.5, 1]$). As an example, for $m = 0, 0.3, 0.5$ and $0.57$, Sandri's algorithm has provided respectively the following LCEs $(0, -0.10, -0.12, -202.14)$, $(+0.25, 0, -1.5, -212.24)$, $(+1.23, 0, -1.2, -207.09)$ and $(0, 0, -0.026, -206.2)$. Then, following the works of Klein and Baier [1991], a classification of (autonomous) continuous-time attractors of dynamical system (\ref{eq13}) on the basis of their Lyapunov spectrum, together with their Hausdorff dimension is presented in Tab. 2. LCEs values have been also computed with the Lyapunov Exponents Toolbox (LET) developed by Pr. Steve Siu for MatLab$^{\mbox{\scriptsize{\textregistered}}}$ and involving the two algorithms proposed by Wolf \textit{et al.} [1985] and Eckmann and Ruelle [1985] (see https://fr.mathworks.com/matlabcentral/fileexchange/233-let). Results obtained by both algorithms are consistent.

\begin{table}[h]
\centering
\caption{Lyapunov characteristics exponents of dynamical system (\ref{eq13}) for various values of $m$.}
{\begin{tabular}{c c c c}\\[-2pt]
\toprule
{\hspace{11mm} $m$} & LCE spectrum & Dynamics of the attractor & Hausdorff dimension  \\[6pt]
\hline\\[-2pt]
{\hspace{3.5mm} $m=0$} & ($ 0, -, -, - $) & Periodic Motion (Limit Cycle) & $D = 1$ \\[1pt]
{$0 < m < 0.5$} & ($ +, 0, -, - $) & 2-Chaos & $D = 3.0007$  \\[2pt]
{$0.5 < m < 1$} & ($ 0, 0, -, - $) & 3-Torus & $D = 2$  \\[2pt]
\botrule
\end{tabular}}
\label{tab2}
\end{table}

\section{Comparison with the UJT circuit}

\subsection{Torus breakdown}

The electronic circuit built in our laboratory (see Fig. 6) has enables to highlight various oscillatory regimes of the UJT according to the value of the control parameter $m$. So, in order to show the ability of our model (\ref{eq13}) to reproduce the experimentally observed phenomena, let's compare the oscilloscope snapshots of the UJT circuit for some values of $m$ and the corresponding phase portraits (see Fig. 9 \& 10). Let's notice that in Fig. 9-b \& 10 we have plotted experimental real time Poincar\'{e} sections which have been obtained by plotting a maximum of voltage signal versus the successive one and which provide evidence of torus breakdown.

\begin{figure}[htbp]
  \begin{center}
    \begin{tabular}{ccc}
      \includegraphics[width=7cm,height=7cm]{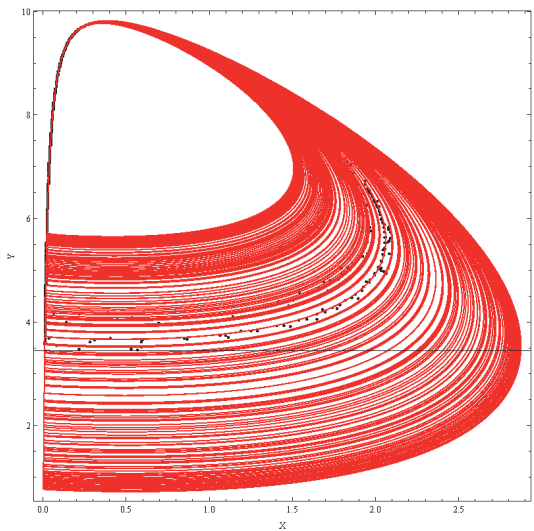} & ~~~~~~ &
      \includegraphics[width=8.5cm,height=7cm]{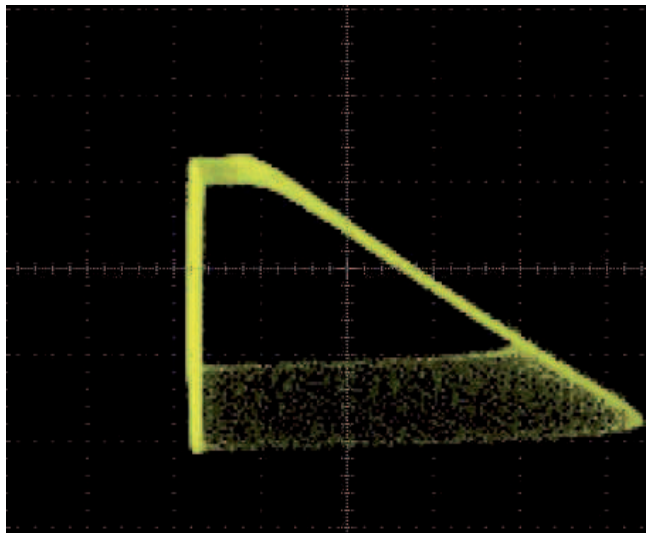} \vspace{0.1in} \\
      (a) $m = 0.51$ & & (b) $m = 0.51$ \\[0.2cm]
    \end{tabular}
    \caption{Phase portrait of dynamical system (\ref{eq13}) and oscilloscope snapshot of UJT circuit for $m=0.51$.}
    \label{fig9}
  \end{center}
  \vspace{-0.5cm}
\end{figure}

\begin{figure}[htbp]
  \begin{center}
    \begin{tabular}{ccc}
      \includegraphics[width=7.05cm,height=7.05cm]{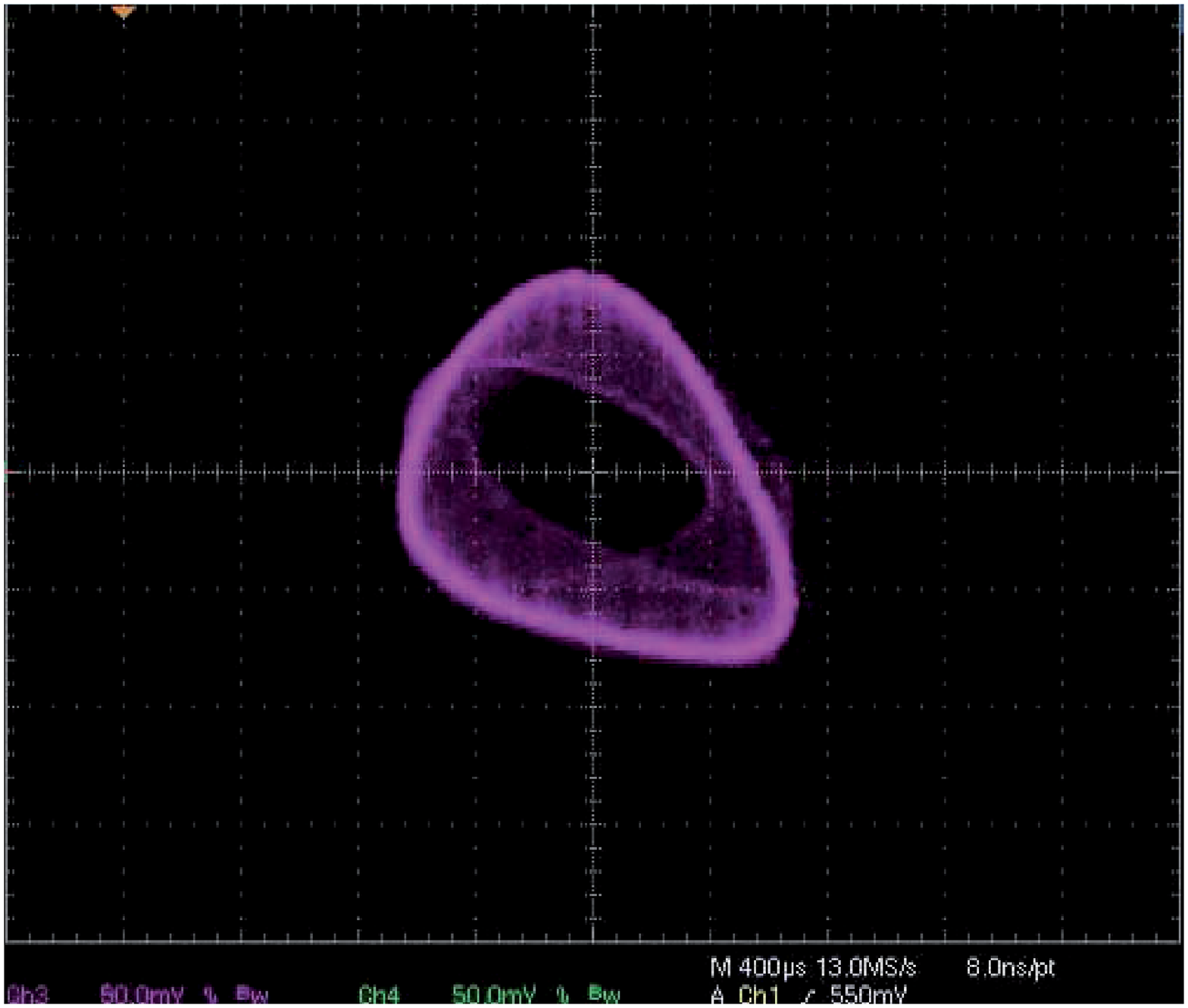} & ~~~~~~ &
      \includegraphics[width=7.05cm,height=7.05cm]{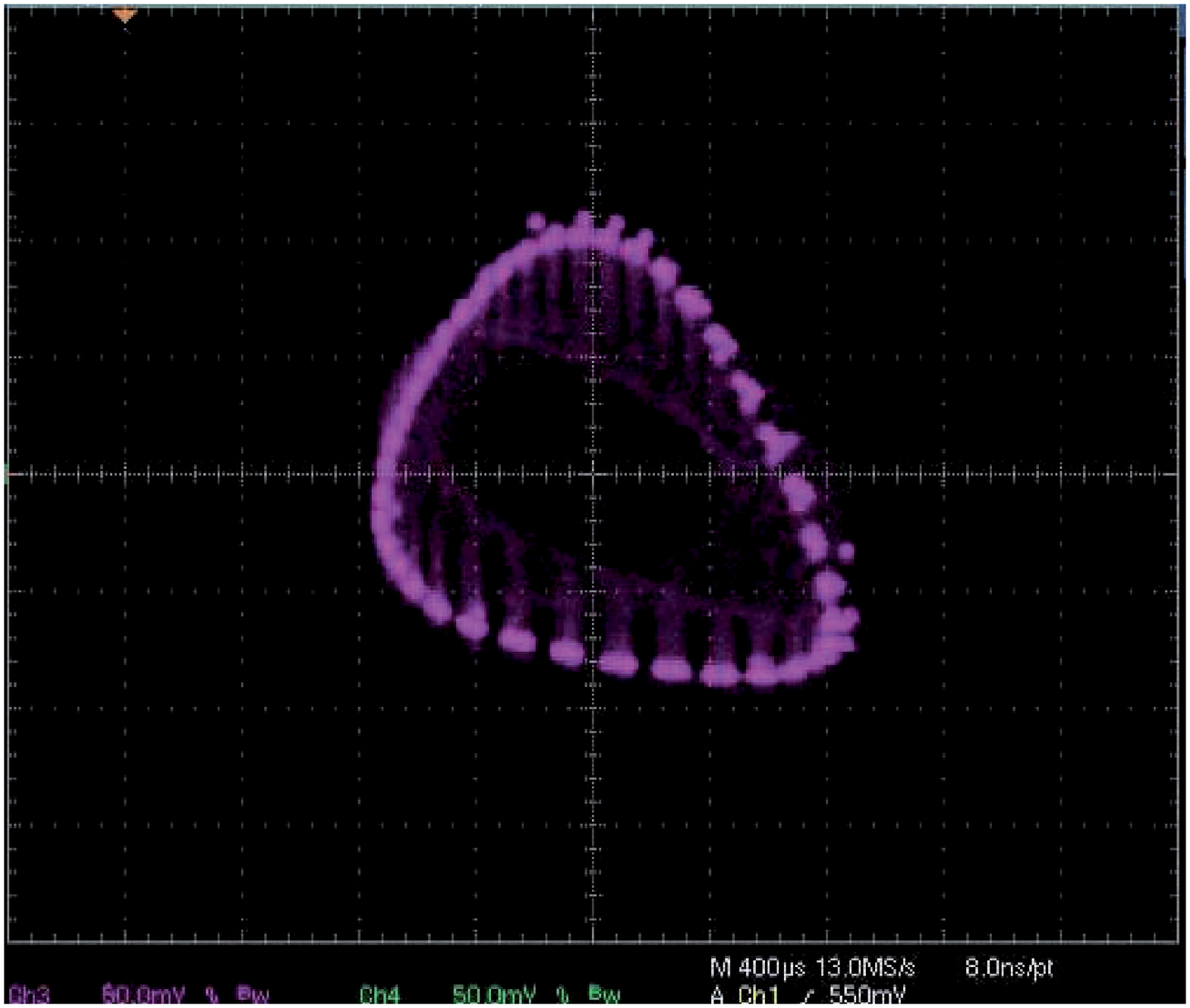} \vspace{0.1in} \\
      (a) $m = 0.39$ & & (b) $m = 0.47$ \\[0.2cm]
      \includegraphics[width=7.05cm,height=7.05cm]{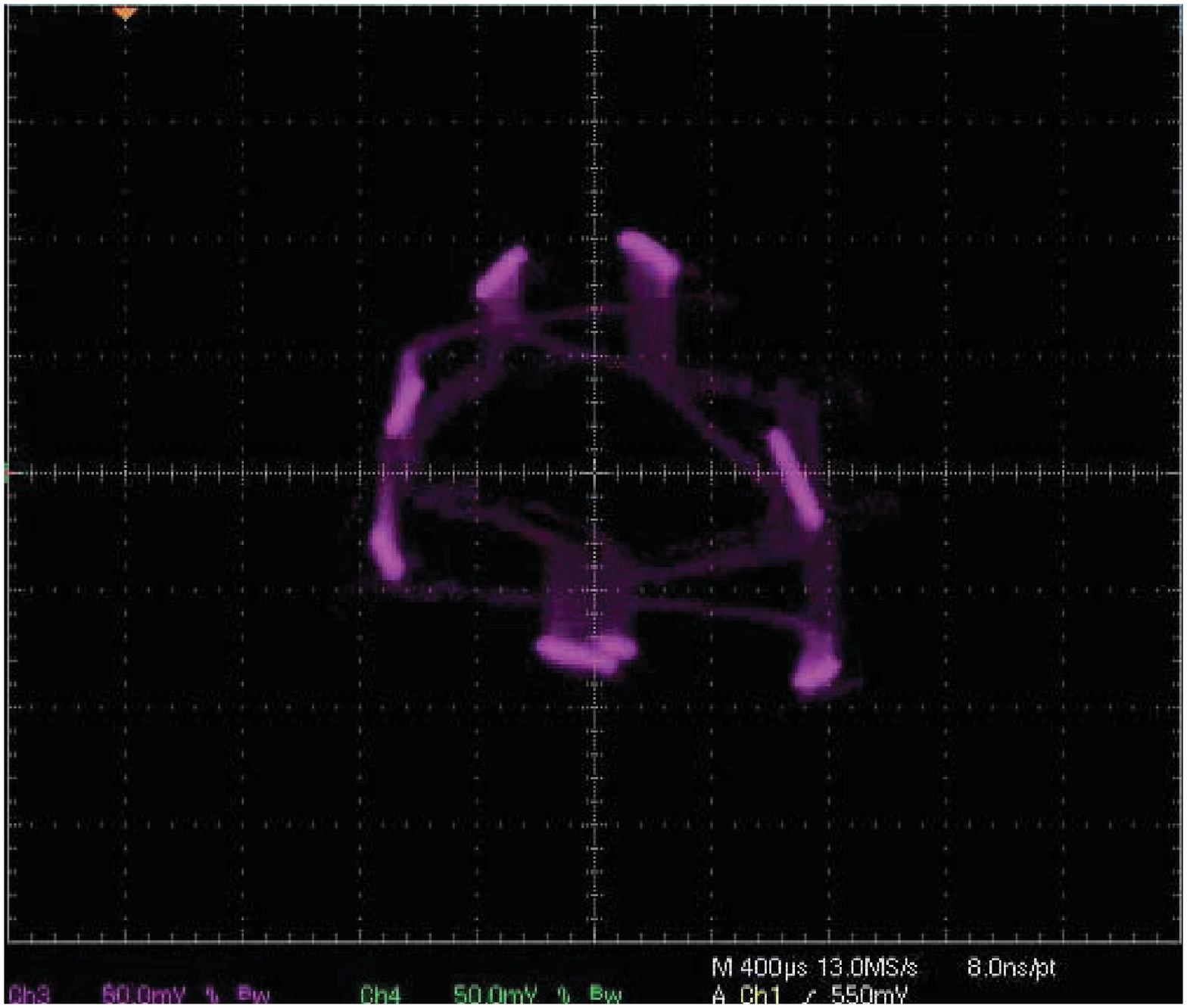} & ~~~ &
      \includegraphics[width=7.05cm,height=7.05cm]{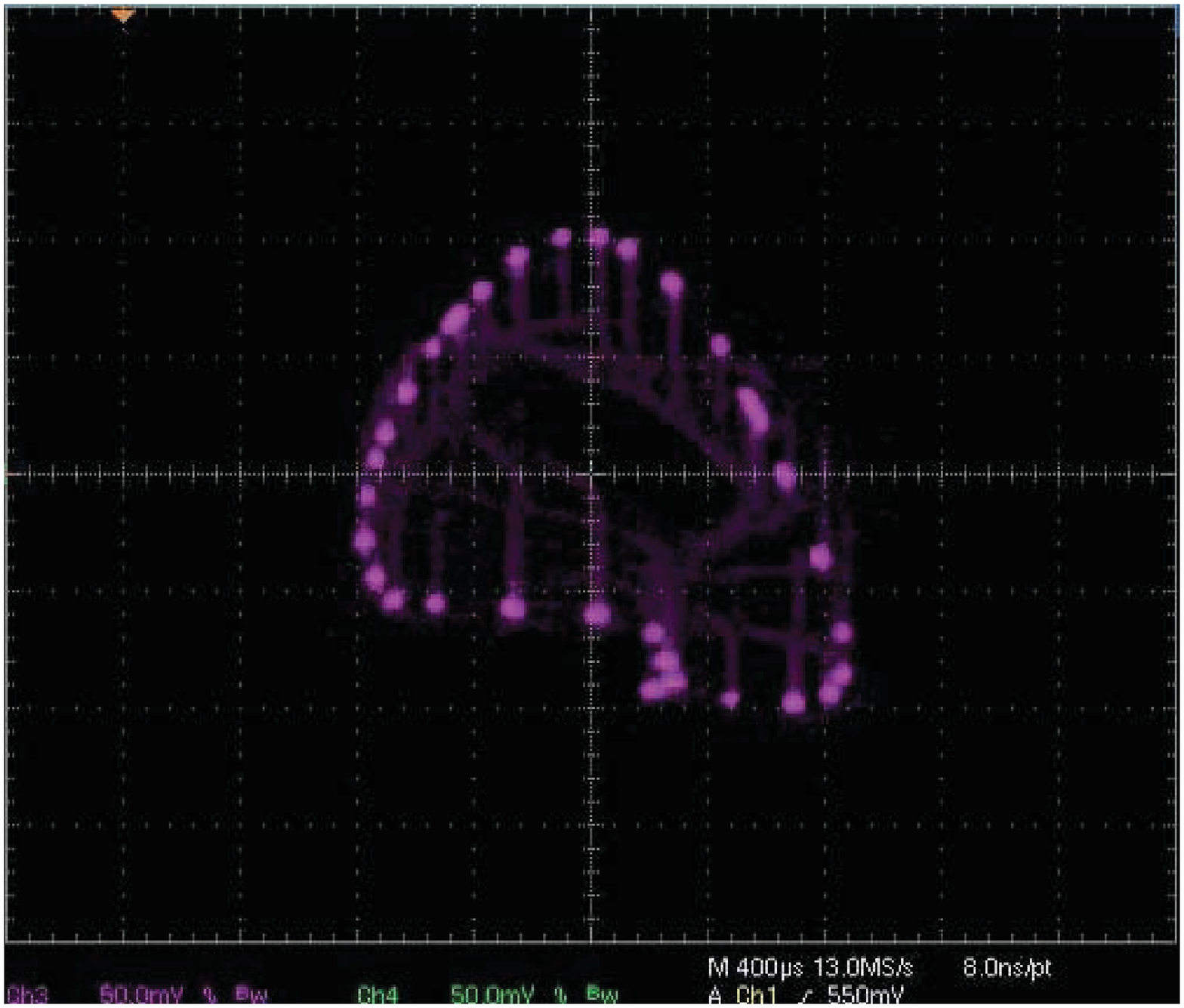} \vspace{0.1in} \\
      (c) $m = 0.49$ & & (d) $m = 0.51$ \\[0.2cm]
      \includegraphics[width=7.05cm,height=7.05cm]{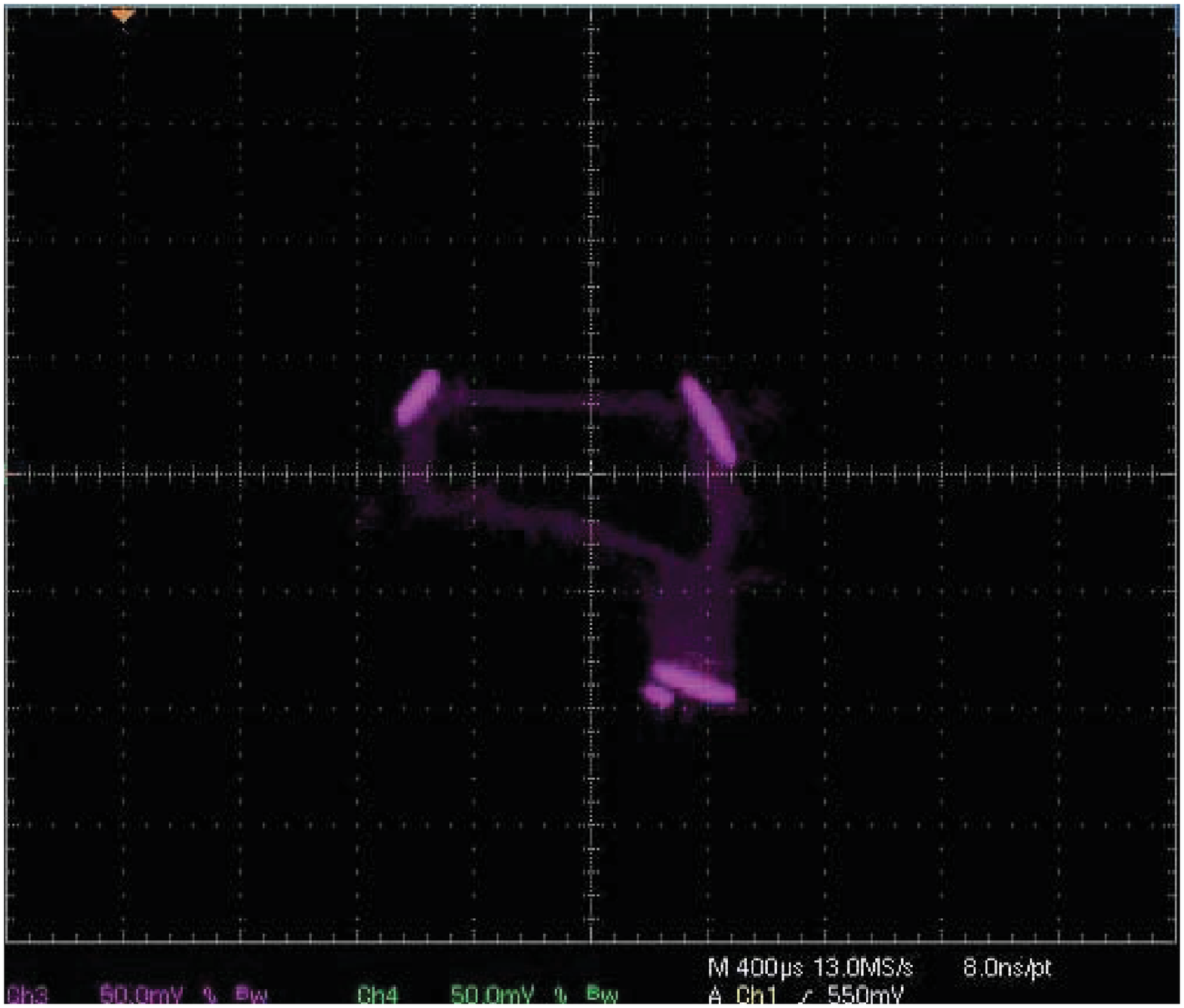} & ~~~ &
      \includegraphics[width=7.05cm,height=7.05cm]{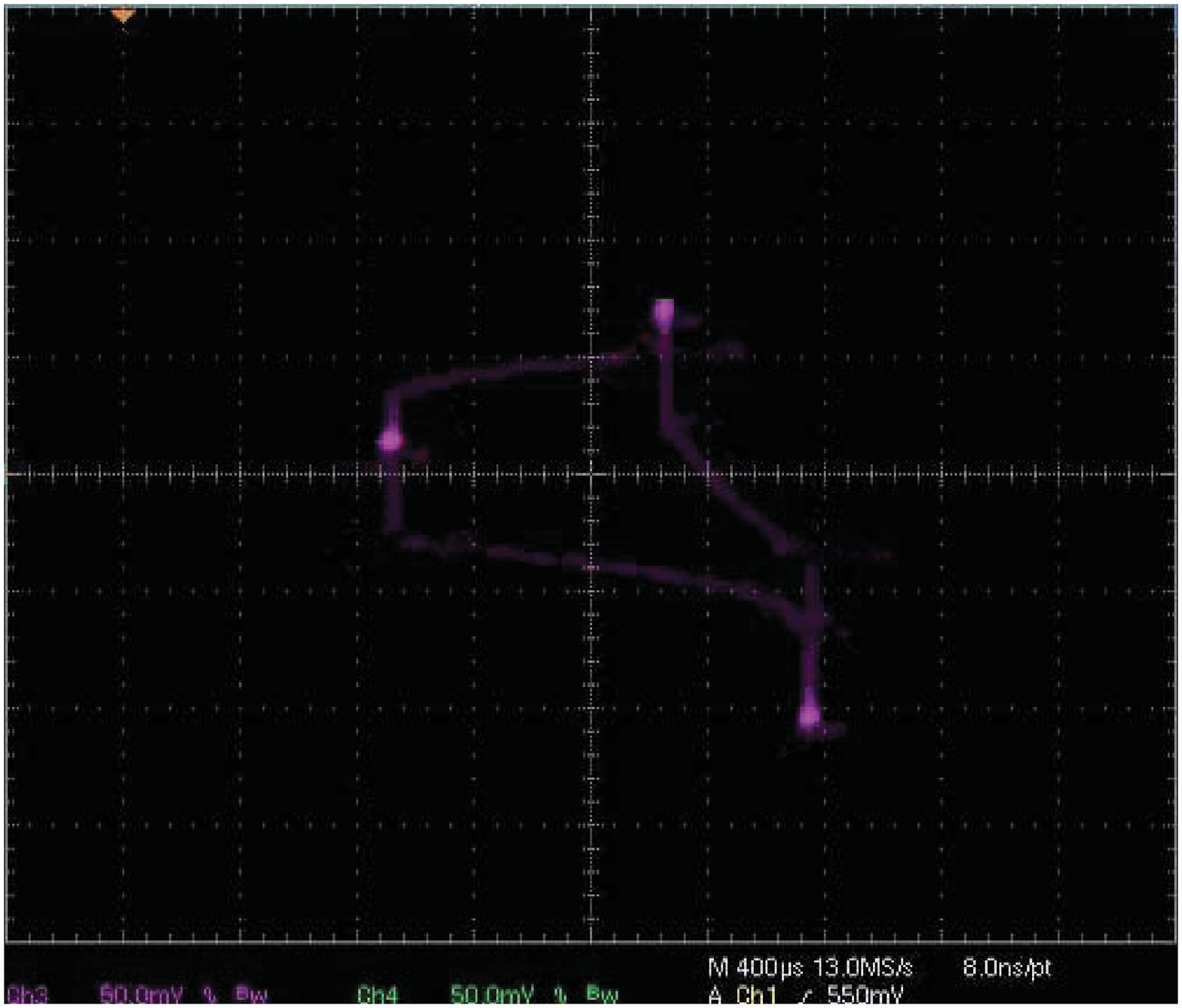} \vspace{0.1in} \\
      (d) $m = 0.514$ & & (e) $m = 0.545$ \\[-0.2cm]
    \end{tabular}
    \caption{Torus breakdown in the UJT circuit for various values $m$.}
    \label{fig10}
  \end{center}
  \vspace{-0.5cm}
\end{figure}

\newpage

\subsection{Pinched hysteresis loop}

According to Leon Chua \cite{Chua2011} ``\textit{pinched hysteresis loop} confined to the first and the third quadrants of the $v-i$ plane whose contour shape in general changes with both the amplitude and frequency of any periodic ``sine-wave-like'' input voltage source, or current source'' is the fingerprint of a memristor or a memristive device. Thus, in order to confirm the memristive feature of the unijunction transistor (UJT) we have applied a sinusoidal signal to the emitter of the UJT and recorded $V_E-I_E$
for many frequencies and amplitudes. The ``sine-wave-like'' input voltage source was always positive and spanned in amplitude from 0 to 3.2 Volts. The results, plotted in Fig. 11-a \& 11-b, show the existence of a ``\textit{pinched hysteresis loop}'' confined to the first quadrant of the $V_E-I_E$ plane.

\begin{figure}[htbp]
  \begin{center}
    \begin{tabular}{ccc}
      \includegraphics[width=6cm,height=6cm]{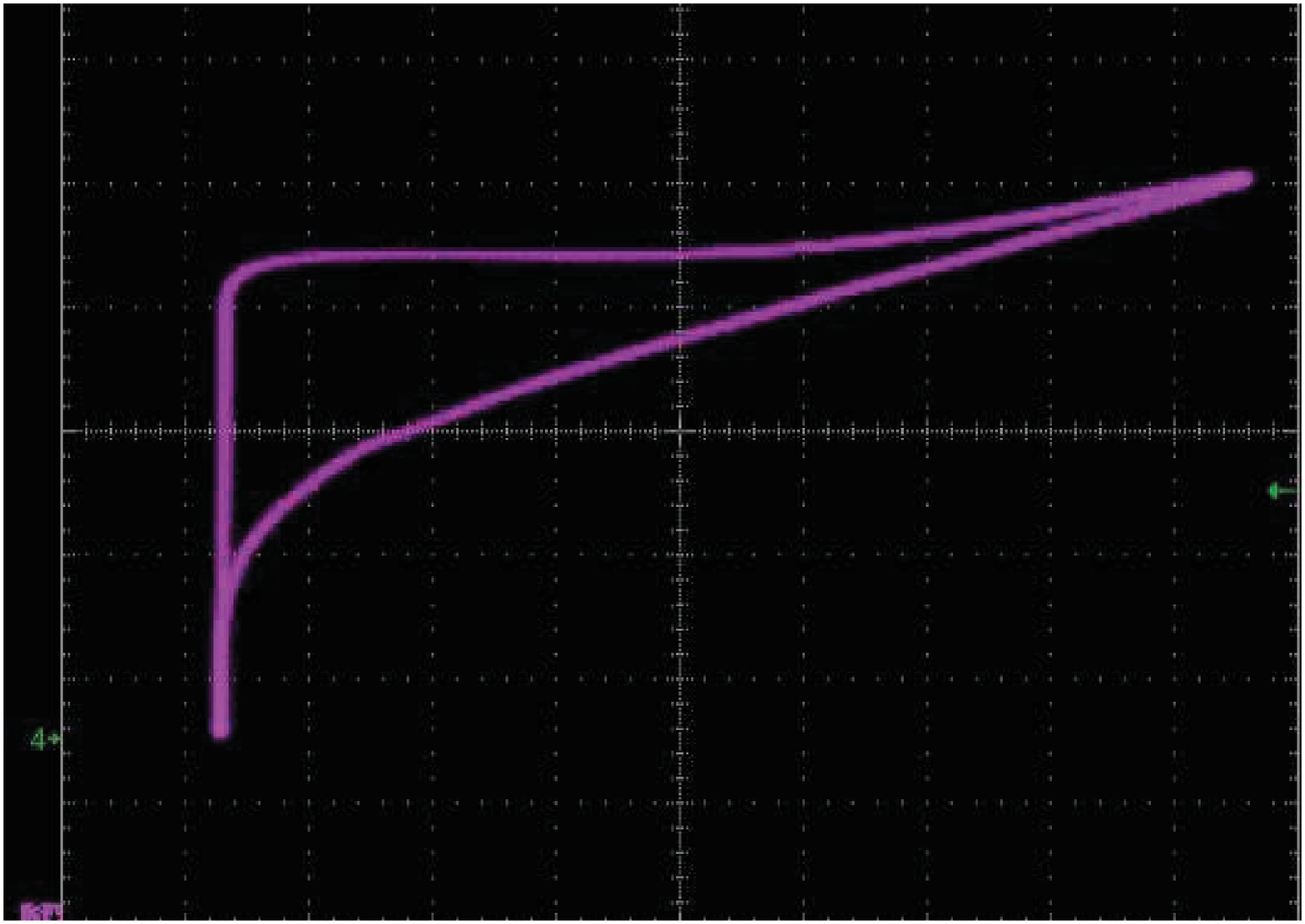} & ~~~~~~ &
      \includegraphics[width=6cm,height=6cm]{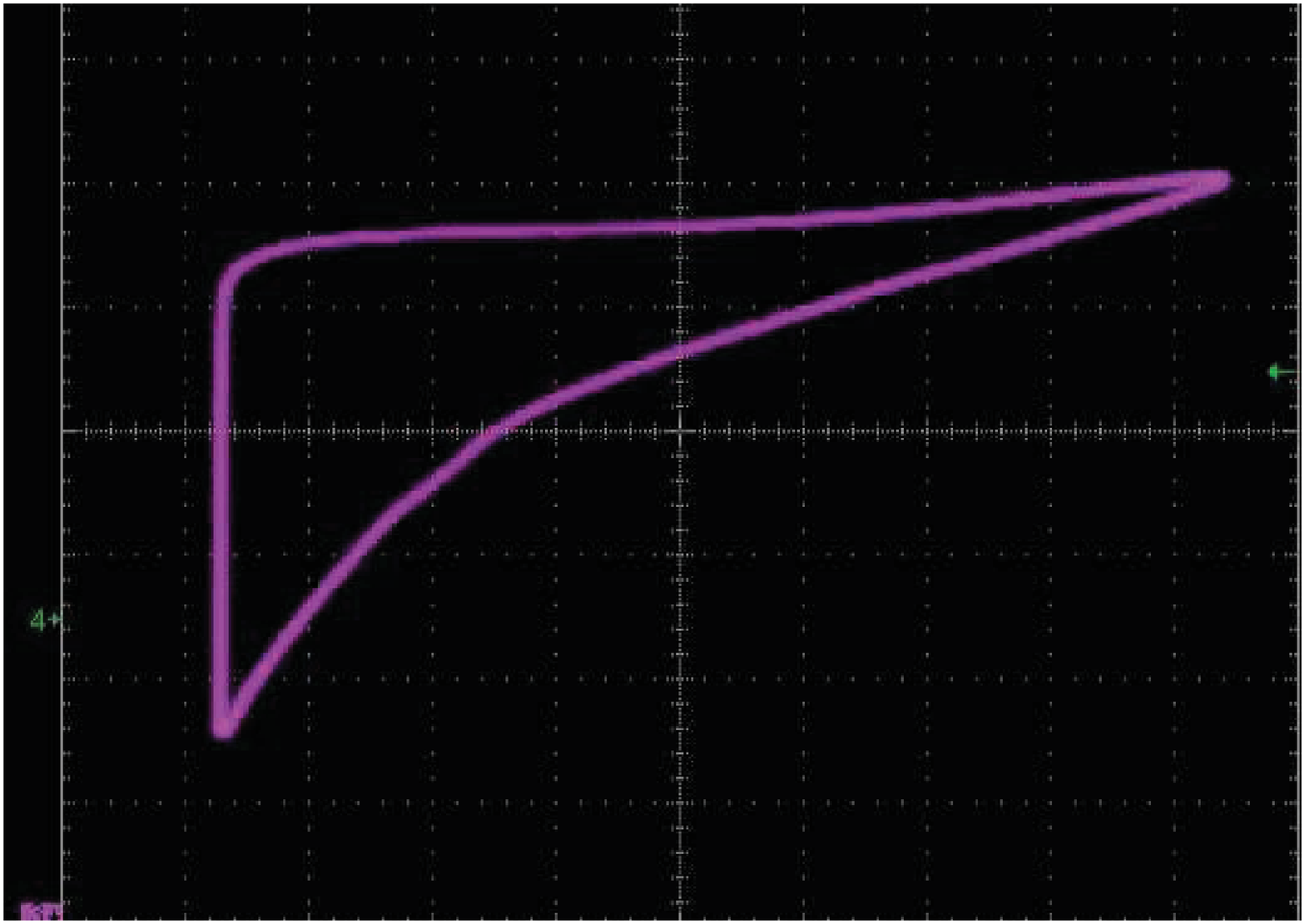} \vspace{0.1in} \\
      (a) $f = 20kHz$ & & (b) $f = 50kHz$ \\[0.2cm]
    \end{tabular}
    \caption{Current-voltage $V_E-I_E$ characteristics of the UJT for a ``sine-wave-like'' input signal of varying frequencies and amplitudes.}
    \label{fig11}
  \end{center}
  \vspace{-0.5cm}
\end{figure}

\section{Discussion}

Starting from a great similarity between the static emitter characteristic of the UJT and the direct current (DC) $v_M-i_M$ static characteristic of the memristor, we proposed to use the memristor's direct current characteristic equation (\ref{eq1}) for modeling the UJT's DC current-voltage characteristic. Then, experimental measurements of the current and voltage of the UJT have indicated a quite good fit of these data for this memristor model (\ref{eq1}), the coefficient of determination being $R^2 = 0.9802$. Thus, this has lead us to suggest to rename the UJT ``Uni Junction Memristor''. By coupling the UJT with an oscillating circuit, we have analyzed the various oscillatory regimes depending on a control parameter $m$ and proposed a new four-dimensional dynamical system allowing to reproduce the experimentally observed phenomena. The bifurcation diagram (Fig. 7) has highlighted a transition from a limit cycle to chaos leading to the appearance of a torus. This scenario has been confirmed by the computation of the Lyapunov characteristic exponents. So, varying the value of the control parameter $m$ from 1 to zero has enabled to highlight the torus breakdown observed in our circuit. Thus, the comparison of our results with the oscilloscope snapshots of the UJT circuit for some values of $m$ has shown that the proposed model was able to reproduce the experimentally observed phenomena, i.e., the occurrence of a torus breakdown (Fig. 8, 9 \& 10). Moreover, the existence of a ``\textit{pinched hysteresis loop}'' confined to the first quadrant of the current-voltage $V_E-I_E$ characteristics of the UJT (Fig. 11) has then confirmed the memristive behavior of the UJT. The importance of this four-dimensional dynamical system is related to a general feature of the coupling between a relaxation oscillator (singing arc, triode, neon tube, UJT, memristor, \ldots) and a driven harmonic oscillator which leads to a torus breakdown.

\section*{Acknowledgments}
Authors would to thank Pr. Tito Arecchi for his helpful advices and comments.


\newpage

\begin{thebibliography}{9}

\bibitem[Bj\"{o}rck(1996)]{Bjorck1996} Bj\"{o}rck, A. [1996] {\it Numerical Methods for Least Squares Problems}, SIAM, Philadelphia, pp. 407.

\bibitem[Chua(1971)]{Chua1971} Chua, L. O. [1971] ``Memristor -- The Missing Circuit Element,'' {\it IEEE Transactions on Circuit Theory} {\bf 18} (5), p.~507-519.

\bibitem[Chua(2011)]{Chua2011} Chua, L. O. [2011] ``Resistance switching memories are memristors,'' {\it Applied Physics A} {\bf 102} (4), p.~765-783.

\bibitem[Eckmann \& Ruelle(1985)]{Eckmann1985} Eckmann, J.P. \& Ruelle, D. [1985] ``Ergodic Theory of Chaos and Strange Attractors,'' {\it Rev. Mod. Phys.}, \textbf{57}, p.~617-656.

\bibitem[Ginoux \& Rossetto(2013)]{Ginoux2013} Ginoux, J.-M. \& Rossetto, B. [2013], ``The Singing Arc: The Oldest Memristor?'' {\it Chaos, CNN, Memristors and Beyond}, in \textit{Chaos, CNN, Memristors and Beyond: A Festschrift for Leon Chua}, World Scientific Publishing, Adamatsky, A. \& Chen, G. (Eds), p.~494-507.

\bibitem[Ginoux {\it et al.}(2017)]{Ginoux2017} Ginoux, J.-M., Meucci, R. \& Euzzor, S., ``Torus Breakdown and Homoclinic Chaos in a Glow Discharge Tube,''
{\it International Journal of Bifurcation and Chaos}, {\bf 27} (14), 1750220.

\bibitem[Klein \& Baier(1991)]{KleinBaier1991} Klein, M., and G. Baier [1991] {\it Hierarchies of dynamical systems}, In {\it A Chaotic Hierarchy}, edited by G. Baier and M. Klein. Singapore: World Scientific.

\bibitem[Lyapunov(1892)]{lyapunov1892} Lyapunov, A.M. [1892] {\it The General Problem of the Stability of Motion} (In Russian), Doctoral dissertation, Univ. Kharkov 1892, English translations: {\it The General Problem of the Stability of Motion}, (A. T. Fuller trans.) Taylor \& Francis, London 1992, pp. 250.

\bibitem[Muthuswamy \& Chua(2010)]{Muthuswamy2010} Muthuswamy, B. \& Chua, L. O. [2010] ``Simplest chaotic circuit,''
{\it International Journal of Bifurcation and Chaos}, Vol. 20, No. 5, p.~1567-1580.

\bibitem[Poincar\'{e}(1892)]{Poin1892} Poincar\'{e}, H. [1892-93-99] {\it Les M\'{e}thodes Nouvelles de la M\'{e}canique C\'{e}leste}, vol. I, II \& III, (Gauthier-Villars, Paris).

\bibitem[Pugliese {\it et al.}(2015)]{Pugliese2015} Pugliese, E., Meucci, R., Euzzor, S., Freire, J.G. \& Gallas, J.A.C. [2015] ``Complex dynamics of a dc
glow discharge tube: Experimental modeling and stability diagrams,'' {\it Sci. Rep.}, \textbf{5}, 08447.

\bibitem[Sandri(1996)]{Sandri1996} Sandri M. [1996] ``Numerical Calculation of Lyapunov Exponents,'' {\it The Mathematica Journal}, \textbf{6}(3), p.~78-84.

\bibitem[Stathopoulos {it et al.}(2017)]{Statho2017} Stathopoulos, S., Khiat, A., Trapatseli, M., Cortese, S., Serb, A., Valov, I. \& Prodromakis, T. [2017] ``Multibit memory operation of metal-oxide bi-layer memristors,'' {\it Scientific Reports}, (13 December 2107) 7, 17532.

\bibitem[Strukhov {\it et al.}(2008)]{Strukhov2008} Strukhov, D. B., Snider, G. S., Stewart, G. R. \& Williams R. S. [2008] ``The missing memristor found,'' {\it Nature}, {\bf 453}, p.~80-83.

\bibitem[Ward(2005)]{Ward2005} Ward, J., ``An Oral History of Jerry Suran,'' \url{http://www.semiconductormuseum.com/Transistors/GE/OralHistories/Suran/Interview.htm}.

\bibitem[Wolf, Swift, Swinney \& Vastano(1985)]{Wolf1985} Wolf, A., Swift, J.B., Swinney, H.L. \& Vastano, J.A. [1985] ``Determining Lyapunov Exponents from a Time Series,'' {\it Physica D}, \textbf{16}, p.~285-317.

\end{thebibliography}
\end{document}